\documentclass[pra,twocolumn,showpacs]{revtex4}
\usepackage{amsmath,amssymb,mathrsfs}
\usepackage{psfrag}
\usepackage{graphicx}
\usepackage{graphics}
\usepackage{epsfig}
\usepackage{bm}
\usepackage{color}
\usepackage{verbatim,color,ulem}

\newcommand{\beq}{\begin{equation}}
\newcommand{\eeq}{\end{equation}}
\newcommand{\beqa}{\begin{eqnarray}}
\newcommand{\eeqa}{\end{eqnarray}}

\renewcommand\Re{\operatorname{Re}}
\renewcommand\Im{\operatorname{Im}}

\begin{document}

\title{Beliaev damping of the Goldstone mode in atomic Fermi superfluids}
\author{G. Bighin$^{1,2}$, L. Salasnich$^{1,3}$, P.A. Marchetti$^{1,2}$ and F. Toigo$^{1,3}$}
\affiliation{$^{1}$Dipartimento di Fisica e Astronomia ``Galileo Galilei'', 
Universit\`a di Padova, Via Marzolo 8, 35131 Padova, Italy
\\
$^{2}$Istituto Nazionale di Fisica Nucleare, Sezione di Padova, 
Via Marzolo 8, 35131 Padova, Italy
\\
$^{3}$Consorzio Nazionale Interuniversitario per le Scienze Fisiche 
della Materia (CNISM), Unit\`a di Padova, Via Marzolo 8, 35131 Padova, Italy
}
\date{\today}

\begin{abstract}

Beliaev damping in a superfluid is the decay of a collective excitation into two lower frequency collective excitations; it represents the only decay mode for a bosonic collective excitation in a superfluid at $T=0$. The standard treatment for this decay assumes a  linear spectrum, which in turn implies that the final state momenta must be collinear to the initial state. We extend this treatment, showing that the inclusion of a gradient term in the Hamiltonian yields a realistic spectrum for the bosonic excitations; we then derive a formula for the decay rate of such excitations, and show that even moderate nonlinearities in the spectrum can yield substantial deviations from the standard result. We apply our result to an attractive Fermi gas in the BCS-BEC crossover: here the low-energy bosonic collective excitations are density oscillations driven by the phase of the pairing order field. These collective excitations, which are gapless modes as a consequence of the Goldstone mechanism, have a spectrum which is well established both theoretically and experimentally, and whose linewidth, we show, is determined at low temperatures by the Beliaev decay mechanism.

\end{abstract}


\maketitle

\section{Introduction}

The discovery of superfluidity below the $\lambda$-point in liquid $^4$He \cite{kapitza,allen} provided a stunning demonstration of quantum properties of matter at a macroscopic level, paving the way for the experimental realization, in more recent times, of  condensates of atomic Bose \cite{anderson} and Fermi \cite{regal} gases and of quasiparticles, like in the recently observed exciton-polariton condensate \cite{kasprzak} in a semiconductor microcavity.

Even if very different in nature and in their physical features, these phenomena are essentially a manifestation of the Bose-Einstein condensation of interacting bosons (respectively single atoms for  $^4$He, Cooper pairs/bosonic dimers for ultracold Fermi gases, polaritons in an exciton-polariton condensate); thus from a theoretical point of view it would seem compelling to describe them, at least to some extent, within a common framework.

Indeed, excitations in a superfluid can be described using the quantum hydrodynamics approach developed by Landau \cite{landau}; a clear advantage of this formalism is the possibility of describing superfluids with non-contact interactions and with a varying number of particles by introducing higher order terms by means of a perturbative expansion around the mean field solution.

Collective excitations in a superfluid are destroyed either by Landau damping, due to their interaction with the thermal cloud, or by Beliaev damping, due to their decay into two, or more, lower energy excitations. There is competition between these two damping modes: whereas Landau damping is relevant at finite temperatures, with a vanishing cross section as the temperature goes to zero, Beliaev damping remains the only allowed decay mode at $T=0$.

Therefore the Beliaev decay represents a test of Landau's hydrodynamic theory. First evidences of a phonon decay have been observed in superfluid liquid $^4$He \cite{maris,millis}; more recently the Beliaev decay has been observed in a trapped Bose-Einstein condensate (BEC) of rubidium atoms \cite{hodby,katz}; an analogous process has been proposed in order to explain the absence of equilibrium in one dimensional interacting bosons, see \cite{ristivojevic} and references therein.

In the present paper we focus on Beliaev decay and  derive an improved version of the classical result \cite{beliaev,landau,stringari} based on the observation that while the original derivation requires a nonlinear term in the spectrum, nonetheless it treats the kinematics in a low-momentum approximation as if the spectrum was effectively linear. We show that this treatment can be extended and that, in particular, the inclusion of a gradient term in the Hamiltonian yields a Bogoliubov-like spectrum for the bosonic excitations \cite{salasnich2}. We calculate the decay rate for the Beliaev damping and show that even for low momenta and small nonlinearities a realistic spectrum can give appreciable differences with respect to the linear approximation of the standard result.

The result is applied to an attractive Fermi gas: as the attractive interaction between atoms is tuned, the gas at $T=0$ goes with continuity from a Bardeen-Cooper-Schrieffer (BCS) weakly-interacting regime, to a strongly interacting gas of bosonic dimers. This scenario can be described \cite{nagaosa,altland,stoof} by introducing the complex Cooper pairing field, which will acquire a non-zero expectation value below the critical temperature. As the phase of the order parameter is macroscopically locked below the critical temperature \cite{goldstone,nambu}, spontaneously breaking the $U(1)$ symmetry, its fluctuations correspond to the gapless mode predicted by the Goldstone theorem. These collective modes turn out to be fundamental in quantitatively describing the dynamics of an ultracold Fermi gas \cite{strinati}; after briefly analyzing the Goldstone mode, we show that its linewidth gets substantially enhanced due to the Beliaev decay process. We also show that our improved description of the decay yields substantial deviations from the standard approximation.

\section{Beliaev damping: an improved treatment}

We briefly introduce Landau's quantum hydrodynamics \cite{landau,stringari}, a semi-phenomenological description of a superfluid which can be, however, rigorously justified and derived from the microscopical theory as discussed in \cite{popov}. An exact expression for the internal energy of a classical liquid in a sound wave is $ E=\int \mathrm{d}^3 x (\frac{1}{2} \rho v^2 + \rho e) $, where $v$ is the local velocity of the fluid, and $\rho$ the local density. Here $e$ represents the internal energy of the fluid for unit mass; Landau's original treatment \cite{landau,beliaev} assumes it to be dependent only on the density $\rho$, and as a consequence the dispersion relation for the sound waves is linear. On the other hand by adding a gradient term as
\beq
e(\rho) \rightarrow e(\rho, \nabla \rho) = e(\rho) + \lambda \frac{\hbar^2}{8m^2} \frac{(\nabla \rho)^2}{\rho^2}
\eeq
higher order terms appear in the dispersion relation as shown in \cite{salasnich2}, $m$ being the mass of a fluid particle, $\lambda$ being a dimensionless coefficient which can be fixed {\it a posteriori}. Within the quantum hydrodynamics framework the velocity and density fields of a fluid are promoted to quantum operators, so that the Hamiltonian for a quantum fluid is:
\beq
\hat{H} = \int \mathrm{d}^3 x \left[ \frac{1}{2} \hat{\mathbf{v}} \cdot \hat{\rho} \hat{\mathbf{v}} + \hat{\rho} e(\hat{\rho}, \nabla \hat{\rho}) \right]
\label{eq:hamiltonian}
\eeq
where the term involving the velocity operator has been opportunely symmetrized to be Hermitean. We rewrite the velocity in terms of a velocity potential $\hat{\mathbf{v}} = \nabla \hat{\phi}$ and the density by separating the equilibrium value $\rho$ from its fluctuations as $ \hat{\rho} = \rho + \hat{\rho}' $. The new operators can be written expanding in plane waves:
\beq
\hat{\rho}' = \frac{1}{\sqrt{2 V}} \sum_{\left| \mathbf{k} \right| \neq 0} A_\mathbf{k} \left( \hat{b}_\mathbf{k}  e^{\mathbf{i} \mathbf{k} \cdot \mathbf{r}} + \hat{b}^\dagger_\mathbf{k}  e^{- \mathbf{i} \mathbf{k} \cdot \mathbf{r}} \right)
\label{eq:rhodef}
\eeq
\beq
\hat{\phi} = \frac{1}{\sqrt{2 V}} \sum_{\left| \mathbf{k} \right| \neq 0} \mathrm{i} \hbar B_\mathbf{k} \left( b_\mathbf{k}  e^{\mathbf{i} \mathbf{k} \cdot \mathbf{r}} - b^\dagger_\mathbf{k}  e^{- \mathbf{i} \mathbf{k} \cdot \mathbf{r}} \right)
\label{eq:phidef}
\eeq
with $V$ the volume of the system; the $b_\mathbf{q}$ ($b^\dagger_\mathbf{q}$) operators annihilate (create) a bosonic excitation over the fundamental state of the liquid $|\Omega\rangle$, and obey the canonical commutation relationships.

We impose that $\hat{\rho}'$ and $\hat{\phi}$ should be canonically conjugate variables
\beq
[ \hat{\phi} (\mathbf{r}), \hat{\rho}' (\mathbf{r}') ] = - \mathrm{i} \hbar \delta (\mathbf{r} - \mathbf{r}')
\eeq
this constraint being fulfilled by $ B_\mathbf{k} = A_\mathbf{k}^{-1}$. The exact treatment of a quantum liquid in Eq. (\ref{eq:hamiltonian}) can be expanded in powers of the field operators: the first to give a contribution is the second order, here the theory can be diagonalized to a theory of non-interacting bosons, i.e. $ \hat{H}^{(2)} = \sum_\mathbf{k} \hbar \omega_\mathbf{k} \hat{b}^\dagger_\mathbf{k} \hat{b}_\mathbf{k}$, and the requirement for $\hat{H}^{(2)}$ to be diagonal fully fixes $A_\mathbf{k}$ as:
\beq
A_\mathbf{k} = \sqrt{\frac{\hbar k \rho}{u}} \left(1 + \lambda \frac{\hbar^2}{8m^2} \frac{k^2}{c^2} \right)^{-\frac{1}{4}}
\eeq
and the dispersion for the bosons has the usual Bogoliubov structure
\beq
\omega_\mathbf{k} = u \hbar k \sqrt{1 + \lambda \frac{\hbar^2}{4m^2} \frac{k^2}{u^2}}
\label{eq:lambdaspectrum}
\eeq
$u$ being the sound velocity of the sound waves in the quantum liquid. Clearly the original linear theory can be recovered by setting $\lambda=0$ and removing the gradient terms. The present formalism, as opposed to the Gross-Pitaevskii equation \cite{gross,pitaevskii}, allows for the decay of a collective excitation in a superfluid, in particular extending the treatment to the third order one immediately sees that the decay of one excitation into two is allowed: this process is the Beliaev decay \cite{beliaev} described above. The third order term of the Hamiltonian is:
\begin{equation}
\begin{split}
\hat{H}^{(3)} = \int \mathrm{d}^3 r \Big[ & ( \nabla \hat{\phi} ) \frac{\hat{\rho}'}{2} ( \nabla \hat{\phi} ) + \frac{1}{6} \left( \frac{\mathrm{d}}{\mathrm{d} \rho} \frac{u^2}{\rho} \right) \hat{\rho}'^3 + \\
& - \lambda \frac{\hbar^2}{8m^2} (\nabla \hat{\rho}')^2 \frac{\hat{\rho}'}{\rho^2} \Big]
\end{split}
\end{equation}
Before going on with the treatment of the Beliaev decay we briefly comment on the scope of application of the present theory; as already mentioned it can be shown \cite{popov} that the hydrodynamic Hamiltonian in Eq. (\ref{eq:hamiltonian}) can be rigorously derived from a description of the Bose gas; this procedure involves integrating out the ``fast fields'', effectively defining a momentum scale $k_c$ below which the perturbative expansion should be valid. 
Following \cite{popov} one can estimate this quantity for a weakly interacting Bose gas; here $k_c$ is the momentum marking the separation between a linear spectrum and the free-particle quadratic spectrum, and from Eq. (\ref{eq:lambdaspectrum}) one gets
\beq
\hbar k_c \simeq \frac{2 m u}{\sqrt{\lambda}}
\label{eq:k0}
\eeq
this condition marking, as argued in \cite{popov}, the upper limit for the validity of the perturbation theory.

In order to study the Beliaev decay we calculate the matrix element:
\beq
H^{(3)}_{if} = \langle i | H^{(3)} | f \rangle
\label{eq:matrixelement}
\eeq
between the following initial and final states:
\beqa
| i \rangle & = & \hat{b}^\dagger_{\mathbf{p}} | \Omega \rangle \\
| f \rangle & = & \hat{b}^\dagger_{\mathbf{q}_1} \hat{b}^\dagger_{\mathbf{q}_2} | \Omega \rangle
\eeqa

The matrix element in Eq. (\ref{eq:matrixelement}), when Eqs. (\ref{eq:rhodef}) and (\ref{eq:phidef}) are plugged in, is essentially the expectation value over $| \Omega \rangle$ of a number of terms composed of six creation/annihilation operators; after a lengthy but straightforward calculation, one obtains
\begin{widetext}
\beq
H^{(3)}_{fi} = \frac{(2 \pi \hbar)^3}{(2V)^\frac{3}{2}} \cdot \delta^{(3)} ( \mathbf{p} - \mathbf{q}_1 - \mathbf{q}_2 ) \cdot 3 \sqrt{\frac{u}{\rho} p q_1 | \mathbf{p} - \mathbf{q}_1|} \left( 1 + \chi_\theta \frac{\rho^2}{u^2} \frac{\mathrm{d}}{\mathrm{d} \rho} \frac{u^2}{\rho} \right)
\label{eq:h3if}
\eeq
\end{widetext}
where $\theta$ is the angle between $\mathbf{p}$ and $\mathbf{q}_1$, the other angles being fixed by the condition $\mathbf{q}_2 = \mathbf{p}-\mathbf{q}_1$ enforced by the $\delta$ function. We also defined:
\beq
\chi^{-1}_\theta = \frac{p-q_1}{\left| \mathbf{p} - \mathbf{q}_1 \right|} (1 + \cos (\theta)) + \cos ( \theta )
\eeq
In deriving Eq. (\ref{eq:h3if}) we neglected all the terms containing $\lambda$; it can be checked that they give $\propto p^7$ and $\propto p^9$ contributions to the decay width, whereas the leading contribution will turn out to be $\propto p^5$. However the nonlinear dispersion relation is relevant when discussing the kinematics: the differential decay rate is calculated using Fermi's golden rule \footnote{The square of the $\delta$ function imposing momentum conservation is to be interpreted as in \cite{landau}: $ \left[ \delta^{(3)} \left( \mathbf{p} - \mathbf{q}_1 - \mathbf{q}_2 \right) \right]^2 = \frac{V}{\left( 2 \pi \hbar \right)^3} \delta^{(3)} \left( \mathbf{p} - \mathbf{q}_1 - \mathbf{q}_2 \right)$}:
\beq
\mathrm{d} w = \frac{2 \pi}{\hbar} | H_{fi}^{(3)} |^2 \delta (E_f - E_i) \frac{V^2}{\left( 2 \pi \hbar \right)^6} \mathrm{d}^3 q_1 \mathrm{d}^3 q_2
\label{eq:decayrate}
\eeq
and $E_f - E_i = \omega_\mathbf{p} - \omega_{\mathbf{q}_1} - \omega_{\mathbf{q}_2}$, $\omega_\mathbf{k}$ is the spectrum as derived in Eq. (\ref{eq:lambdaspectrum}). The integration over $\mathrm{d}^3 q_2$ is performed using the momentum conservation constraint appearing in $| H_{fi}^{(3)} |$, the integration over the angular part of $\mathrm{d}^3 q_1$ removes the $\delta$ function related to energy conservation, fixing at the same time the decay angle $\theta_0$, and finally the radial integration remains explicit. The final result for the decay rate is:
\beq
w = \frac{9}{32 \pi \rho \hbar^4} \int_0^p q^2 | \mathbf{p} - \mathbf{q} |^2_0 \frac{\left( 1 + \chi_{\theta_0} \frac{\rho}{u^2} \frac{\mathrm{d}}{\mathrm{d} \rho} \frac{u^2}{\rho} \right)^2}{| f'(\cos \theta_0, p, q) |} \mathrm{d} q
\label{eq:w}
\eeq
where $| \mathbf{p} - \mathbf{q} |^2_0 = | p^2 + q^2 - 2 p q \cos \theta_0|$ for shortness sake, $f (\cos \theta, p, q)= \frac{1}{u} \frac{|\mathbf{p} - \mathbf{q}|}{p q} \left( \omega_p - \omega_q - \omega_{|\mathbf{p} - \mathbf{q}|}\right)$ is essentially the energy conservation constraint, $f'$ is its derivative with respect to the first argument and $\theta_0 = \theta_0 (p,q)$ is the only zero of $f$ in the interval $[-\pi,\pi]$, and represents the allowed decay angle given the incoming and outgoing momenta.

Equation (\ref{eq:w}) is the main original result of the present paper, which we will apply to an attractive Fermi gas. We stress that $w$ in Eq. (\ref{eq:w}) is a function of just $\rho$, $u$ and of the incoming momentum $p$; moreover the exact form of the spectrum, including the $\lambda$ coefficient, contributes indirectly to the final result, by modifying $f$ and, consequently, $\theta_0$. We also note the kinematics constraints can be satisfied and the decay is allowed only if the aforementioned zero of $f$ exists, an equivalent condition being that the spectrum should grow faster than linearly.

Let us make the physical meaning of the last remark clearer, expanding the spectrum in Eq. (\ref{eq:lambdaspectrum}) in powers of $k$:
\beq
\omega_\mathbf{k} = u k + \alpha k^3 + O(k^4)
\eeq
where $\alpha$ has the same sign as $\lambda$. The energy conservation constraint in the low momentum limit reads $1 - \cos \theta = 3 \alpha (p-q)^2$ and can be fulfilled only if $\alpha \geq 0$, i.e. only if the spectrum grows linearly or more than linearly; for $\alpha < 0$ no decay is allowed.

We now focus on the strictly linear case $\alpha=0$. Energy and momentum can be conserved only if $\theta_0 = 0$, i.e. the momenta of the decaying excitation and those of the decay products are parallel. Even for very small values of $\alpha$ the decay kinematics deviates significantly from the aforementioned linear situation as $\theta_0$ increases.

We stress that, even if the standard treatment of Beliaev decay \cite{landau,beliaev} correctly identifies $\alpha \geq 0$ as a necessary condition for the decay to happen, then 
its use of $\alpha=0$ in the kinematics is a critical assumption; 
on the other hand the present treatment by including the gradient term in Eq. (\ref{eq:hamiltonian}) allows for a realistic, Bogoliubov-like spectrum. 

Let us derive the standard result from the more general Eq. (\ref{eq:w}): having set $\lambda=0$ for a linear spectrum $\omega_\mathbf{k} = u | \mathbf{k} | $ one has that $\theta_0 =0$, $f' = 1$ and also $\chi = \frac{1}{3}$. Moreover noting that $\int_0^p q^2 | p - q |^2 \mathrm{d} q= p^5 / 30$, we recover Beliaev's original approximation \cite{beliaev,landau}, which we report here for the sake of completeness:
\beq
w = p^5 \frac{3}{320 \pi \rho \hbar^4} \left( 1 + \frac{\rho^2}{3 u^2} \frac{\mathrm{d}}{\mathrm{d} \rho} \frac{u^2}{\rho} \right)^2
\label{eq:wx}
\eeq

To conclude the present section we note that in the case of a weakly-interacting Bose gas Eq. (\ref{eq:wx}) further simplifies, because in this case
\beq
\frac{\rho^2}{u^2} \frac{\mathrm{d}}{\mathrm{d} \rho} \frac{u^2}{\rho} = 0
\label{eq:pval}
\eeq
Alternatively the weakly-interacting Bose gas regime can also be investigated, as done in \cite{katz}, starting from the atomic Hamiltonian, introducing the Bogoliubov approximation and isolating the relevant decay vertices. The present hydrodynamic approach is different because it can be derived, as already mentioned, by separating the fast and slow components of the fields, introducing a momentum scale $k_c$, whereas the Bogoliubov approximation merely separates the zero-momentum contribution. We expect the two approaches to yield the same results for $k \lesssim k_c$, as we verified. The hydrodynamic approach, however, is better suited for analyzing the collective excitations of an attractive Fermi gas.

\section{Beliaev damping for an attractive Fermi gas}

\begin{figure*}[th!]
\begin{center}
{\includegraphics[trim=0 0 0 0,width=17.4cm,clip]{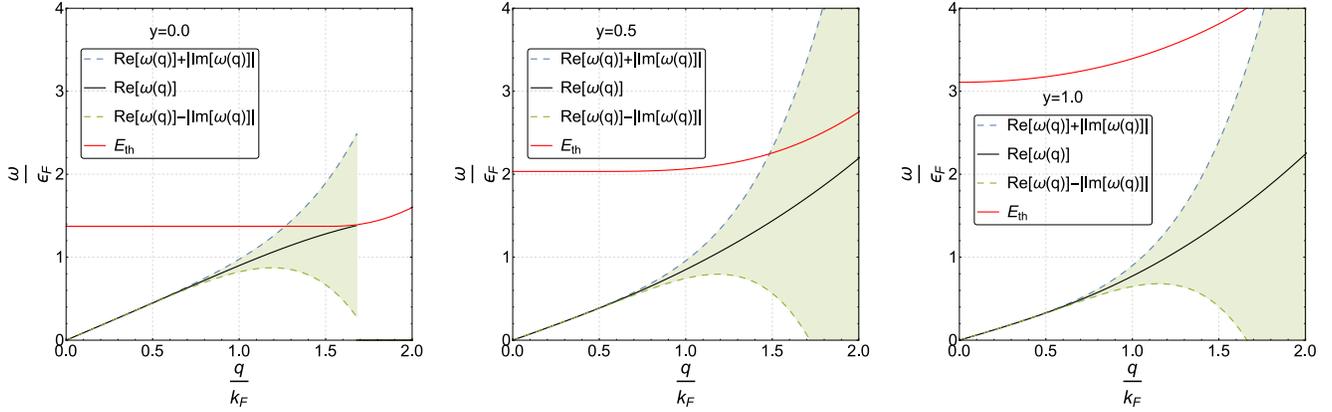}}
\end{center}
\caption{The collective mode spectrum for $y=0.0$, $y=0.5$, $y=1.0$, from left to right. The black line represent the real part of the spectrum, while the dashed lines represent $\pm$ the imaginary part, the bold red line represents the threshold energy $E_{th}$.}
\label{fig:2}
\end{figure*}

Let us consider a three-dimensional, uniform dilute gas of interacting Fermi atoms; the atoms are neutral and have two spin species. This system can be described within the path integral formalism \cite{nagaosa,altland,stoof} in which the fermions are represented by the complex Grassman fields $\psi_\sigma (\mathbf{r}, \tau), \bar{\psi}_\sigma (\mathbf{r}, \tau)$, with the spin index $\sigma=\uparrow, \downarrow$. The partition function for the system at temperature $T$, with chemical potential $\mu$ is:
\beq 
{\cal Z} = \int {\cal D}[\psi_\sigma,\bar{\psi}_\sigma] 
\ \exp{\left\{ -{1\over \hbar} \ S  \right\} } \; , 
\eeq
with the following action and (Euclidean) Lagrangian density:
\beq 
S = \int_0^{\hbar\beta} 
d\tau \int_{V} d^3{\bf r} \ \mathscr{L}
\eeq
\beq 
\mathscr{L} = \sum_\sigma \bar{\psi}_{\sigma} \left[ \hbar \partial_{\tau} 
- \frac{\hbar^2}{2m}\nabla^2 - \mu \right] \psi_{\sigma} 
+ g \, \bar{\psi}_{\uparrow} \, \bar{\psi}_{\downarrow} 
\, \psi_{\downarrow} \, \psi_{\uparrow} 
\eeq
as usual $\beta=1/(k_B T)$, $k_B$ is the Boltzmann constant, $V$ is the volume of the system and $g<0$ is the strength of the contact interaction between atoms; this quartic interaction can be decoupled, as usual, through a Hubbard-Stratonovich transformation in the Cooper channel, introducing the pairing field $\Delta (\mathbf{r}, \tau) \sim \psi_\downarrow \psi_\uparrow$. Now the Euclidean Lagrangian density reads:
\beqa 
\mathscr{L} =
\sum_\sigma \bar{\psi}_\sigma \left[  \hbar \partial_{\tau} 
- {\hbar^2\over 2m}\nabla^2 - \mu \right] \psi_\sigma + \nonumber \\
+ \bar{\Delta} \, \psi_{\downarrow} \, \psi_{\uparrow} 
+ \Delta \bar{\psi}_{\uparrow} \, \bar{\psi}_{\downarrow} 
- {|\Delta|^2\over g}
\label{ltilde}
\eeqa
In order to obtain the partition function $\mathcal{Z}$ one also has to extend the functional integration to the newly introduced pairing fields $\Delta$, $\bar{\Delta}$.
We rewrite the fields $\Delta$, $\bar{\Delta}$ as the sum of their saddle point value plus the fluctuations
\beq
\Delta({\bf r},\tau) = \Delta_0 +\eta({\bf r},\tau)
\label{eq:fluctuations}
\eeq
Up to this point the theory is exact; the mean field theory of a Fermi gas is simply found by neglecting the fluctuations $\eta(\mathbf{r}, \tau)$, $\bar{\eta}(\mathbf{r}, \tau)$. The functional integration defining the partition function $\mathcal{Z}$ can then be performed, yielding:
\beq 
{\cal Z}_{mf} = \exp{\left( - \beta \, \Omega_{mf} \right)}
\eeq
with:
\beq 
\Omega_{mf} = -\sum_{\bf k} \left( E_{sp}(\mathbf{k}) - \epsilon_\mathbf{k} + \mu \right) 
- V {\Delta_0^2\over g} \; . 
\label{omega0-div}
\eeq
with $\epsilon_\mathbf{k}=\hbar^2k^2/(2m)$; $E_{sp}(\mathbf{k})=\sqrt{(\epsilon_\mathbf{k}-\mu)^2+\Delta_0^2}$ is the spectrum of elementary single-particle fermionic excitations. The number and the gap equations for the system can be readily obtained from the mean-field grand potential $\Omega_{mf}$:
\beqa
N = - \frac{\partial \Omega_{mf}}{\partial \mu} \\
\left( \frac{\partial \Omega_{mf}}{\partial \Delta_0} \right)_\mu = 0
\label{eq:neqgeq}
\eeqa
Lastly the gap equation in Eq. (\ref{eq:neqgeq}) needs to be regularized, and this can be done by replacing $g$ with the scattering length $a_s$, according to the following prescription (see e.g. \cite{leggett}):
\beq
\frac{m}{4 \pi a_s} = - \frac{1}{g} + \frac{1}{V} \sum_{\mathbf{k}} \frac{1}{2 \epsilon_\mathbf{k}}
\eeq
where $a_s$ is the s-wave scattering length associated to the interatomic potential.

\begin{figure}[th!]
\begin{center}
{\includegraphics[trim=0 0 0 0,width=8.7cm,clip]{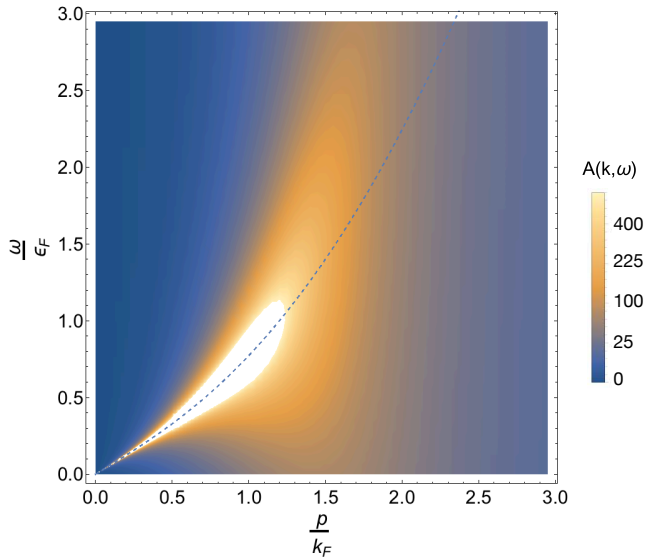}}
\end{center}
\caption{The pair fluctuation spectral function $A_{\eta \eta}(\mathbf{k}, \omega)$ for $y=1$, the dashed line shows the corresponding spectrum. For $\frac{p}{k_f} \gtrsim 1$ the line broadening due to the Beliaev decay effectively destroys a collective excitation, this is also approximately the scale marking the end of the validity of the perturbative approach. For comparison here $E_{th} > 3 \epsilon_F$}
\label{fig:3}
\end{figure}

Let us now analyze the fluctuations contribution to the present theory; going back to Eq. (\ref{eq:fluctuations}) and reinstating the fluctuations fields $\eta (\mathbf{r},\tau)$, $\bar{\eta} (\mathbf{r},\tau)$ up to the quadratic (Gaussian) order \cite{ohashi2,randeria2} the partition function reads:
\beq 
\mathcal{Z} = \mathcal{Z}_{mf} \ \int 
{\cal D}[\eta,\bar{\eta}] \ 
\exp{\left\{ - {S_g(\eta,\bar{\eta}) \over \hbar} \right\}} \; , 
\label{sigo}
\eeq
having defined the Gaussian action:
\beq 
S_{g}(\eta,\bar{\eta}) = {1\over 2} \sum_{q} 
({\bar\eta}(q),\eta(-q)) \ {\mathbb{M}}(q) \left(
\begin{array}{c}
\eta(q) \\ 
{\bar\eta}(-q) 
\end{array}
\right) \; 
\label{eq:mmatrix}
\eeq
with $q=(\mathbf{q}, \mathrm{i} \omega_n)$, and $\omega_n=\frac{2 \pi}{\beta}n$ are the Bose Matsubara frequencies. The matrix in Eq. (\ref{eq:mmatrix}) is the inverse propagator for the pair fluctuations of an interacting Fermi gas, its matrix elements are defined by \cite{randeria2,tempere2}:

\beqa
\mathbb{M}_{11} (q) = \frac{1}{g} + \sum_\mathbf{k} \left( \frac{u^2 u'^2}{\mathrm{i} \omega_n - E - E'} - \frac{v^2 v'^2}{\mathrm{i} \omega_n + E + E'} \right) \\
\mathbb{M}_{12} (q) = \sum_\mathbf{k} u v u' v' \left( \frac{1}{\mathrm{i} \omega_n + E + E'} - \frac{1}{\mathrm{i} \omega_n - E - E'} \right)
\eeqa
where $u=u_{\mathbf{k}} = \sqrt{\frac{1}{2} ( 1 + \frac{\epsilon_\mathbf{k} - \mu}{E_{sp} (\mathbf{k})})}$, $v=v_\mathbf{k}=\sqrt{1-u^2_{\mathbf{k}}}$, $u'=u_{\mathbf{k}+\mathbf{q}}$, $v'=v_{\mathbf{k}+\mathbf{q}}$, $E=E_{sp} (\mathbf{k})$, $E'=E_{sp} (\mathbf{k} + \mathbf{q})$. The remaining matrix elements are defined by the relations: $ \mathbb{M}_{22} (q)  = \mathbb{M}_{11} (-q) $, $ \mathbb{M}_{21} (q)  = \mathbb{M}_{12} (q) $.
By integrating out the $\eta(\mathbf{r},\tau)$, $\bar{\eta}(\mathbf{r},\tau)$ fields we get the Gaussian contribution to the grand potential \cite{randeria2,tempere2}:
\beq
\Omega_g = \frac{1}{2 \beta} \sum_q \ln ( \det \mathbb{M} (q))
\eeq

A completely equivalent description can be given, after a unitary transformation, in terms of the (linearized) phase and amplitude of the fluctuation field, which can be decomposed as $\eta (\tau, \mathbf{x}) = \left( \lambda (\tau, \mathbf{x}) + \mathrm{i} \theta (\tau, \mathbf{x}) \right) / \sqrt{2}$. The Gaussian level action now reads:
\beq
S_g = \frac{1}{2} \sum_\mathbf{q} \begin{pmatrix}
\lambda^* & \theta^* \\
\end{pmatrix} \begin{pmatrix}
\mathbb{M}^E_{11} + \mathbb{M}_{12} & \mathrm{i} \mathbb{M}^O_{11} \\
- \mathrm{i} \mathbb{M}^O_{11} & \mathbb{M}^E_{11} - \mathbb{M}_{12} \\
\end{pmatrix} \begin{pmatrix}
\lambda \\
\theta
\end{pmatrix}
\label{eq:actionmp}
\eeq
in terms of the even/odd components in $\mathrm{i} \omega_n$ of the $\mathbb{M}$ matrix elements \cite{gubankova,engelbrecht}, i.e. $\mathbb{M}_{ab}^{E/O} (\mathbf{q},\mathrm{i} \omega_n)= \frac{1}{2} (\mathbb{M}_{ab} (\mathbf{q},\mathrm{i} \omega_n) \pm \mathbb{M}_{ab} (\mathbf{q},-\mathrm{i} \omega_n))$. This representation makes clear that as soon as the Cooper pairing field $\Delta (\mathbf{r},\tau)$ acquires a non-zero expectation value, i.e. under $T_c$, as a consequence of the $U(1)$ symmetry breaking one expects to observe the gapless Goldstone mode \cite{nagaosa}. More specifically it can be verified from Eq. (\ref{eq:actionmp}) that for $T=0$, in the low momentum limit the phase and amplitude fluctuations are decoupled \cite{engelbrecht}: the diagonal entries in Eq. (\ref{eq:actionmp}) go to zero, and the phase (Goldstone) mode is gapless, while the amplitude (Higgs) mode exhibits a mass gap. From now on up to the end of the present section we will study the system at $T=0$. Focusing on the former mode, we observe that, indeed, by solving for $\omega$ the equation
\beq
\det \mathbb{M} (\mathbf{q}, \mathrm{i} \omega_n \to \omega) = 0
\label{eq:modes}
\eeq
we obtain the spectrum of the bosonic collective mode, showing a gapless branch. Notably in the BEC regime $y \gtrsim 1$, and across the whole crossover for low enough momenta, this mode takes (within very good approximation) the familiar Bogoliubov-like form
\beq 
E_{col}(\mathbf{q}) = \sqrt{\epsilon_\mathbf{q} \left( \lambda \epsilon_\mathbf{q} + 2 m c_s^2 \right)} 
\label{eq:ecol}
\eeq
and the sound speed $c_s$, along with the parameter $\lambda$, depends on $y=1/(k_F a_s)$. We use this spectrum in the deep BEC limit, while in the intermediate regime near unitarity we solve numerically Eq. (\ref{eq:modes}) to get the ``exact'' spectrum within the present Gaussian approximation scheme. When comparing the ``exact'' spectrum so obtained with the Bogoliubov approximate form, one also has to remember that a natural momentum scale can be defined by studying whether and when the dispersion enters the two-particle continuum reaching the threshold energy:
\beq
E_{th} (\mathbf{q}) = \min_\mathbf{k} ( E_\mathbf{k} + E_{\mathbf{k} + \mathbf{q}})
\eeq
above which a Cooper pair breaks down in two fermions. As far as the present work is concerned it is important noting that $E_{col}$ grows more (less) than linearly if $\lambda>0$ ($\lambda<0$), moreover the parameter $\lambda$ can be calculated easily either from a numerical solution of Eq. (\ref{eq:modes}) or using the techniques in \cite{marini2}. It turns out that $\lambda$ is a monotonically increasing function of $y=1/(k_F a_s)$, where $k_F$ is the Fermi momentum. In particular $\lambda$ takes negative values in the deep BCS regime and changes its sign for $y = y_c \approx -0.14$; referring to the previous section we can then conclude that no Beliaev decay will happen in the deep BCS region for $y < y_c$.

We now want to adapt Eq. (\ref{eq:w}) to the present theory. We start by noticing that if the spectrum has the form in Eq. (\ref{eq:ecol}), then the decay angle $\theta_0$ defined in the previous section has an analytic expression:
\begin{widetext}
\beq
\cos \theta_0 (p,q) =\frac{m^2 c_s^2}{\lambda  p q \hbar^2}+\frac{q}{2 p}+\frac{p}{2 q}- \frac{\sqrt{m^2 c_s^4+2 m c_s^2 \lambda \left(\epsilon_p+\epsilon_q \right)-2 \lambda  p q \frac{\hbar^2}{2m}
   \sqrt{\left(2 m c_s^2+\lambda  \epsilon_p\right) \left( 2 m c_s^2+\lambda  \epsilon_q\right)}+
   \lambda ^2 \left(\epsilon^2_p+\epsilon^2_q\right)}}{2 \lambda  p q \frac{\hbar^2}{2m}}
\label{eq:theta0}
\eeq
\end{widetext}
We note that for the special case $\lambda=1$, $2 m c_s^2 = 2$ Eq. (\ref{eq:theta0}) coincides with the result in \cite{katz}.

Finally the more complicated expression inside the parenthesis in Eq. (\ref{eq:w}) can be expressed using the techniques devised in \cite{salasnich} as:
\beq
\frac{\rho^2}{u^2} \frac{\mathrm{d}}{\mathrm{d} \rho} \frac{u^2}{\rho} = -\frac{30 \epsilon (y)-8 y \epsilon '(y)-3 y^2 \epsilon ''(y) + y^3 \epsilon'''(y)}{30 \epsilon (y) -18 y \epsilon '(y) + 3 y^2 \epsilon ''(y)}
\label{eq:pval2}
\eeq
as a function of $\epsilon(y)=\frac{5}{3} \epsilon_F^{-1} \mathcal{E}$, where $\mathcal{E}$ is the bulk energy per particle per particle of an interacting Fermi gas; when calculating our final results we compared $\epsilon(y)$ as fitted in \cite{salasnich} from experimental data with its mean field counterpart, observing no appreciable differences as far as the quantity Eq. (\ref{eq:pval2}) is concerned. Consistently with the result found in Eq. (\ref{eq:pval}) for the weakly-interacting Bose gas, the quantity in Eq. (\ref{eq:pval2}) tends to zero in the deep BEC limit.

\begin{figure}[hb!]
\begin{center}
{\includegraphics[trim=0 0 0 0,width=8.2cm,clip]{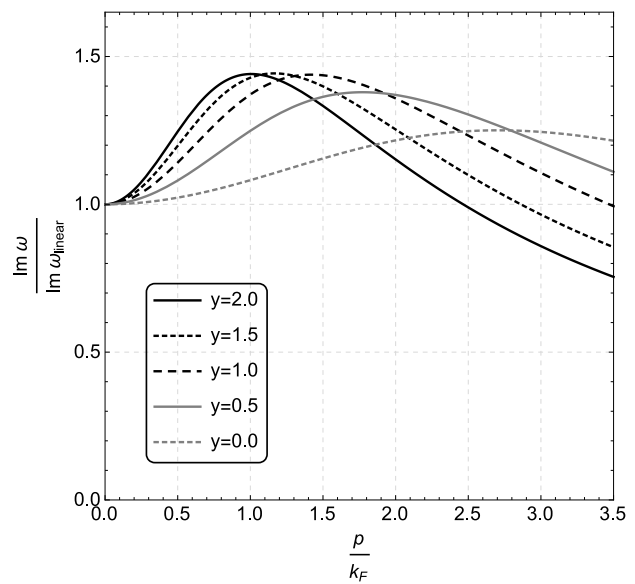}}
\end{center}
\caption{The Beliaev decay width calculated from Eq. (\ref{eq:w}) divided by the original Beliaev result (linear approximation), for different values of $y=1/(k_F a_s)$.}
\label{fig:4}
\end{figure}

We calculate the Beliaev decay width for the Goldstone collective mode of an attractive Fermi gas; as previously noted there is no decay in the BCS regime up to $y=y_c \approx -0.14$, as the spectrum as a function of $|\mathbf{q}|$ grows less than linearly. For higher values of $y$ we can associate an imaginary part to the Goldstone mode spectrum, as
\beq
\Im \omega_\mathbf{p} = - \frac{\hbar w}{2}
\eeq
using $w$ from Eq. (\ref{eq:w}). In Fig. (\ref{fig:2}) we report the real spectra $\omega_\mathbf{p}$, obtained from Eq. (\ref{eq:modes}), for three values of $y=1/(k_F a_s)$, from unitarity to the BEC regime ($y=0$, $y=0.5$, $y=1$), along with their imaginary part due to the Beliaev decay.

A collective excitation in a superfluid Fermi gas cannot have arbitrarily high energy, as it will be damped either by the dissociation mechanism at the threshold energy $E_{th}$, decaying into two fermions, or by the Beliaev mechanism, decaying into two lower frequency collective excitations. Either way a natural energy cutoff can be associated to a bosonic excitation.

Referring to the left pane of Fig. (\ref{fig:2}) we start at unitarity ($y=0$) where the Beliaev decay width is quite narrow: here a collective excitation will mainly decay by hitting the threshold energy $E_{th}$ and breaking down into two fermions \cite{combescot}. On the other hand, approaching the BEC regime ($y=0.5$, $y=1.0$) the Beliaev decay width gets larger before the collective spectrum touches $E_{th}$: here the preferred decay mode for a collective excitation will be decaying into two lower frequency collective excitations. This trend, i.e. the progressively bigger importance of the Beliaev mechanism approaching the BEC regime, can be observed by comparing the three panes in Fig. (\ref{fig:2}).

In order to define an energy cutoff due to the Beliaev mechanism, we can match the real and the imaginary part of $\omega_\mathbf{p}$ similarly identifying a scale beyond which a collective excitation is no longer well-defined due to the Beliaev decay. This remark is made clear by looking at the pair fluctuation spectral function
\beq
A_{\eta \eta}(\mathbf{k}, \omega) = - 2 \Im G_{\eta \eta} (\mathbf{k},\omega + \gamma_\mathbf{k})
\label{eq:a}
\eeq
plotted in Fig. (\ref{fig:3}). As noted in \cite{klimin}, it can be interpreted as the contribution to the density from the fluctuations at a given wave number $\mathbf{q}$ and a given momentum $\omega$. In the previous equation $\omega$ is assumed to be real and $\gamma_\mathbf{k}=-\frac{\hbar w}{2}$ is the imaginary component of the spectrum due to the Beliaev decay, $G_{\eta \eta}$ is the Green's function obtained by inverting the matrix in Eq. (\ref{eq:mmatrix}) and taking the $(1,1)$ entry. We observe that for low momenta most of the spectral weight is peaked around the dispersion relation, which is marked by a dashed line, assuming the usual Lorentzian structure. However, as the spectrum continues after $p \simeq k_F$, for high momenta the line broadening effect due to the Beliaev decay effectively destroys the collective excitation, and the spectral weight is distributed over a large region. The border between these two regimes can also be approximately found by imposing the aforementioned condition $\Re \omega_\mathbf{k} = \Im \omega_\mathbf{k}$, which can be easily read from Fig. (\ref{fig:3}): when the real part of the dispersion is bigger than the imaginary part, the expression in Eq. (\ref{eq:a}) has a narrow peak; as the imaginary part of the spectrum gets bigger the Lorentzian structure of the peak is lost and the excitation is no longer well defined.

We can conclude that, as we go from the BCS to the BEC regime, the dissociation mechanism at $E_{th}$ gets less and less relevant, as the collective mode spectrum gets further away from $E_{th}$; at the same time, the Beliaev decay channel opens at $y=y_c$ and gets progressively more relevant. Finally in Fig. (\ref{fig:4}) we compare the decay width, as predicted by the present theory, with the original linear approximation \cite{beliaev,landau}: even for relatively small momenta our treatment shows relevant differences with respect to the standard treatment. The differences get larger in the BEC regime, consistently with the fact that the nonlinearity term $\lambda$ in the spectrum is bigger; however we stress that even for nearly linear spectra, see the cases $y=0$ and $y=0.5$ in Fig. (\ref{fig:2}), the correction due to the present treatment can amount up to 25\% for $\frac{p}{k_F} \simeq 1$.

In conclusion we briefly comment on the scope of applicability of the present theory to the fermionic case; adapting Eq. (\ref{eq:k0}) one finds
\beq
k_c \simeq \frac{2 m c_s}{\sqrt{\lambda}}
\eeq
and we do not expect the theory to be applicable above this momentum threshold; a direct calculation shows that starting at unitarity, up to the moderate BEC regime we considered in Fig. (\ref{fig:2}) and Fig. (\ref{fig:3}), $k_c$ assumes respectively the following values: $k_c=3.06 k_F$, $k_c=1.72 k_F$, $k_c=1.31 k_F$; in the deep BEC limit we get
\beq
k_c \simeq {1\over 2}{\Delta_0^2\over |\mu|}
\eeq
We notice that  for the cases we considered, the momentum scale $k_c$ marking the breakdown of the perturbation theory is higher or equal to the scale at which a collective excitation is no longer well defined due to a high decay rate; we conclude then that the present treatment is consistent.

\section{Conclusions}

We have extended the standard result of Landau's hydrodynamic theory of a superfluid, which leads to a purely linear spectrum implying a collinear Beliaev decay. By including a gradient term in the Hamiltonian, we have recovered the Bogoliubov-like spectrum bosonic excitations in a superfluid have a Bogoliubov-like spectrum of excitations producing a larger phase-space for the Beliaev. We have shown that even slight variations from linearity of the spectrum can give important modifications to the decay rate of the process we consider.

We have applied our result to an interacting Fermi gas in the BCS-BEC crossover: we have shown that no decay happens at zero temperature in the deep BCS regime, due to kinematics constraints, as the spectrum grows less than linearly. As the strength of the attractive interaction is increased, the collective mode spectrum increases linearly or faster as $y \gtrsim -0.14$, thus allowing the decay of one collective excitation  into lower energy excitations and this mechanism becomes more and more relevant as the coupling gets stronger. We observe that in the BCS regime in the low-temperature limit a collective excitation can decay only by breaking down into two fermions at $E_{th}$; on the other hand at unitary and in the BEC regime a collective excitation can also decay in two collective excitations by means of the Beliaev decay.

Finally we have identified the regimes to which the theory we have developed applies. The perturbation theory behind the hydrodynamic treatment of a quantum liquid breaks down at a critical momentum $k_c$. We have estimated this value, verifiying the internal consistence of our treatment, showing by analyzing the decay width and the pair fluctuation spectral function that in the fermionic case $k_c$ is higher or equal to the momentum scale at which a collective excitation is no longer well defined due to the decay process.

\begin{acknowledgments}
Work partially supported by MIUR (Ministero Istruzione Universit\`a e Ricerca) through PRIN Project ``Collective Quantum Phenomena: from Strongly-Correlated Systems to Quantum Simulators".
\end{acknowledgments}

\end{document}